\documentclass[twocolumn,aps,prc,epsf]{revtex4}

\usepackage{amsfonts}
\usepackage{amsmath}
\usepackage{amssymb}
\usepackage{graphicx}

\newcommand{\trD}[1]{\mbox{\boldmath $#1$}}

\newcommand{\vers}[1]{\hat{\trD{#1}}}
\newcommand{\spu}[2]{u_{#1}(#2)}
\newcommand{\spudag}[2]{u^\dag_{#1}(#2)}
\newcommand{\spub}[2]{\overline{u}_{#1}(#2)}
\newcommand{\spv}[2]{v_{#1}(#2)}
\newcommand{\spvdag}[2]{v^\dag_{#1}(#2)}

\begin{document}

\title{Chiral Symmetry Breaking and Scalar Confinement}
\author{P. Bicudo}
\email{bicudo@ist.utl.pt}
\author{G. M. Marques}
\email{gmarques@cfif.ist.utl.pt}
\affiliation{Dep. F\'{\i}sica and CFIF, Instituto Superior T\'ecnico, Av.
Rovisco Pais
1049-001 Lisboa, Portugal}
\begin{abstract}
We address the old difficulty in accommodating the scalar
quark-antiquark confining potential together with chiral symmetry
breaking. We develop a quark confining potential inspired in the
QCD scalar flux tube. The coupling to quarks consists in a double
vector vertex. We study the Dirac and spin structure of this
potential. In the limit of massless quarks the quark vertex is
vector. Nevertheless symmetry breaking generates a new scalar
quark vertex. In the heavy quark limit the coupling is mostly
scalar. We solve the mass gap equation and find that this
potential produces spontaneous chiral symmetry breaking for light
quarks. The quantitative results of this model are encouraging.
\end{abstract}
\maketitle

\section{Introduction}
\label{sec:introduction}

Spontaneous Chiral Symmetry Breaking (S$\chi$SB) is accepted to
occur in low energy hadronic physics. Another important feature of
hadronic physics, suggested by the spectroscopy of hadrons, by
lattice simulations and by models of confinement is scalar
confinement. However S$\chi$SB and scalar confinement are
apparently conflicting, since the first requires a chiral
invariant coupling to the quarks, like the vector coupling of QCD.
We address a recent quest of Bjorken \cite{Bjorken}, ``how are the
many disparate methods of describing hadrons which are now in use
related to each other and to first principles of QCD?''. Although
the vector confinement of quarks is not yet ruled out
\cite{Szczepaniak,Lagae,Brambilla,Soto}, here we try to solve this
old chiral symmetry versus scalar confinement conflict of hadronic
physics, which remained open for many years.

In this paper we explore scalar confinement from the perspective
of chiral symmetry breaking. In Section \ref{sec:motivation} we
motivate the importance of both S$\chi$SB and scalar confinement.
We show these features of hadronic physics to have some subtle
weaknesses that we capitalize to construct a model.
The potential used in our Quark Model (QM) is defined in Section
\ref{sec:potential}. The self consistent mass gap equation for the
quarks is derived in Section \ref{sec:mass gap}. In Section
\ref{sec:numerical} we solve numerically the mass gap equation and
calculate the quark condensate. Finally, in Section
\ref{sec:conclusion} we present some conclusions.

\section{Matching chiral symmetry breaking with scalar
confinement} \label{sec:motivation}

The QCD lagrangian is chiral invariant in the limit of vanishing
quark masses. Nambu and Jona-Lasinio showed that including chiral
symmetry in fermionic systems provides a natural explanation for
the small pion mass, which is much lighter than all the other
isovector hadrons. Because of this crucial fact the mechanism of
S$\chi$SB is accepted to occur in low energy hadronic physics for
the light flavors $u$, $d$ and $s$, where $m_u , m_d \ll m_s <
\Lambda_{QCD} < M_N/3$. Similarly to the vector Ward identities in
gauge symmetry, the axial Ward identities constitute a powerful
tool of chiral symmetry. The techniques of current algebra led to
beautifully correct theorems, the Partially Conserved Axial
Current (PCAC) theorems. The different variants of QMs are widely
used as simplifications of QCD. They are convenient to study quark
bound states and hadron scattering. Recently \cite{Bicudo} we have
shown these beautiful PCAC theorems, like the Weinberg theorem for
$\pi-\pi$ scattering, to be reproduced by QMs with S$\chi$SB.
Another important benefit of having S$\chi$SB in the QM is the
reduced number of parameters. The mass gap equation generates a
dynamical constituent quark mass, which is no longer an
independent parameter, even for quarks with a vanishing current
mass. The quark-quark, quark-antiquark, antiquark-antiquark
potentials, and the quark-antiquark annihilation and creation
interactions are all originated in the same chiral invariant
Bethe-Salpeter kernel. Therefore any QM for light quarks should
comply with the S$\chi$SB. Moreover the microscopic coupling of
the quark to the confining interaction should include the vector
coupling which is present in the quark-gluon vertex of QCD.
However there is some evidence that a vector quark-quark potential
is not sufficient to provide the expected scale of S$\chi$SB of
the order of $200-300$ MeV which is present both in the
constituent quark mass and in the quark condensate. It was
realized by Adler \cite{Adler} that a linear confinement with
vector couplings was not sufficient to provide the correct quark
condensate. Moreover the gluon propagator extracted from the
lattice, when used in a one gluon exchange truncation of the quark
mass gap equation, is not able to provide the expected quark
condensate \cite{Bhagwat}. This also happens with the gluon
propagator extracted from the solution of truncated
Schwinger-Dyson equations of QCD \cite{Alkofer}. Importantly,
these gluon propagators exhibit a non-perturbative mass. This mass
should produce a Meissner effect in Yang-Mills fields, and this is
expected to produce a confining string for the quarks. Here we
will estimate the effect of the confining string on the quark
condensate.

On the other hand the confining potential for constituent quarks
is probably scalar. We can learn much by comparing simply the
spectrum of the hydrogen atom with the masses of all hadron
families. In a perturbative QCD scenario, the hadron spectroscopy
would be dominated by the one gluon exchange, which is
qualitatively similar to the one photon exchange interaction that
explains in detail atomic physics. It turns out that all the
hadronic families, say of mesons or baryons, with light or heavy
flavors, show similar differences with the hydrogen spectrum. It
is remarkable that the Spin-Orbit potential (also called fine
interaction in atomic physics) turns out to be suppressed in
hadronic spectra since it is smaller than the Spin-Spin potential
(hyperfine interaction). This constitutes an evidence of
non-perturbative QCD. Another evidence of non-perturbative QCD is
present in the angular and radial excitations of hadrons, which
fit linear trajectories in Regge plots, and suggest a long range,
probably linear, confining potential for the quarks. This led
Henriques, Kellett and Moorhouse \cite{Henriques} to develop a QM
where a short range vector potential and a long range scalar
potential partly cancel the Spin-Orbit contribution. The short
range potential is Coulomb-like (inspired in the one gluon
exchange) and its quark vertex has a vector coupling structure
$\bar \psi \gamma^\mu \psi$. The long range potential has scalar
coupling $\bar \psi \psi$ and is a linear potential
\cite{Henriques,Chan,Isgur,Kwong}.

The same scalar confinement picture is extracted from lattice
simulations. In quenched lattices which simulate the
quark-antiquark potential in the heavy quark limit, the pattern of
spin-spin, tensor, and spin-orbit interactions is compatible with
a scalar confinement \cite{Michael,Gromes,Bali}. Moreover the
presently favored confinement picture in the literature is the
flux tube, or string picture, with tension $\sigma \simeq 200$
MeV/Fm. Quantum mechanics suggests that a thin string, in its
ground-state, should be a scalar object \cite{Allen}. Only higher
energy excitations of the string would have angular momentum. This
was capitalized by Isgur and Paton in the flux tube model
\cite{Isgur2}.

In this paper we will assume the quarks are coupled to a scalar
object which provides a linear confinement. On top of that we want
this coupling to use, as a microscopic building block, the vector
gluon-quark which is present in QCD. Notice that to get a Lorentz
scalar coupling a simple one gluon vertex is not enough. The
coupling needs at least two vertices. The simplest way to achieve
this is to have the string emitting two effective gluons that
couple to the same quark line. This double vertex, coupling the
string to a quark line via two intermediate gluon propagators, is
presented in Fig. \ref{loopvertex}. Effectively this double vertex
is similar to the vertices that couple a quark to a gluon ladder
in the soft pomeron models \cite{Low,Nussinov}. It is also related
to the light quark vertex in the heavy-light quark bound states
studied in the local gauge coordinate
\cite{Simonov,Nora,Nefediev}. Such a double vertex was also
introduced in the coupling of short strings to quarks
\cite{Polikarpov}. In the cumulant expansion formalism int terms
of gluon correlators \cite{Dosh,Simonov} this means the
quark-antiquark coupling will be dominated by four gluon
correlators.

\section{The double vertex non-perturbative confining interaction}
\label{sec:potential}

\begin{figure}
\includegraphics{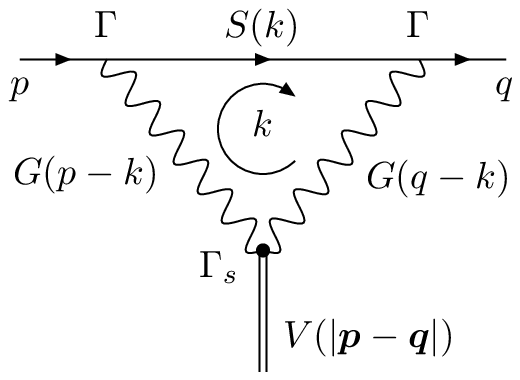}
\caption{The coupling of a quark to a string with a double gluon
vertex} \label{loopvertex}
\end{figure}

In this section we construct the simplest possible coupling, that
simulates that of a quark line with a scalar string using two
vector couplings as building blocks. The most general coupling of
this kind is presented in Fig. \ref{loopvertex} and it is simply
read,
\begin{equation}
\int\frac{d^4k}{(2\pi)^4}(\Gamma V^a)S(k)(\Gamma V^b)G^{bc}(p-k)
(\Gamma_S W^{cde})G^{ad}(q-k) \, ,
\end{equation}
where $\Gamma$ is the dirac structure of the fermion-gluon
interaction and $V^a$ the usual color interaction $\lambda^a/2$.
We denote the quark propagator by $S(k)$ and the gluon propagator
by $G^{ab}(k)$. Finally $\Gamma_S W^{cde}$ represents the coupling
of the gluon pair to the string.

To get the coupling of the string to a light quark, we follow the
coupling obtained in the heavy-light quark system, computed in the
local coordinate gauge \cite{Nora}. This model interpolates
between the heavy-light meson in the local gauge and the effective
QM. So, for the Dirac structure of the quark-gluon sub-vertices
$\Gamma$ we have $\gamma^0$ matrices, which is also compatible
with the Coulomb gauge.

\begin{figure}
\includegraphics{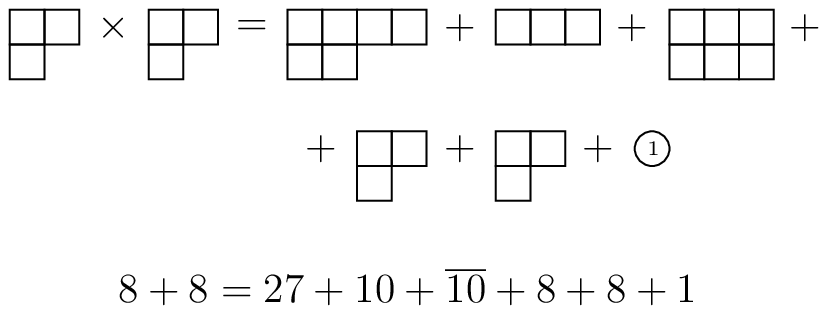}
\caption{String color} \label{String color}
\end{figure}

In the color sector, as already stated, the coupling of the same
sub-vertices has a $\lambda^a /2$ structure, where $\lambda^a$ are
the Gell-Mann matrices. The remaining sub-vertex includes the
coupling of two color octets, see Fig. \ref{String color}. The
string is also a colored object; it contains a flux of
colorelectric field. For a scalar coupling, which is symmetric, we
use the symmetric structure function $d^{abc}$ defined by
\begin{equation}
\{ \lambda^a ,\lambda^b \}= \frac{4}{3}\,\delta^{ab} + 2\,
d^{abc}\lambda^c \ .
\end{equation}

This will result in a color contribution for the effective vertex
of
\begin{equation} \label{color factor}
d^{abc} \, \frac{\lambda^b}{2} \cdot \frac{\lambda^c}{2} = {\cal
C}\, \frac{\lambda^a}{2}, \qquad {\cal C}=\frac{5}{6}\ .
\end{equation}
In QMs the string usually couples to the quark line with a
$\lambda^a /2$. In our case it couples with two $\lambda^a /2$,
one for each sub-vertex. But as we can see from (\ref{color
factor}) the effective result is the same.

The gluon propagators and the different sub-vertices result in a
distribution in the loop momentum $k$. Here different choices
would be possible. For simplicity we assume that the relative
momentum $p-q$ flows equally in the two effective gluon lines. We
also remark that the distribution in $k$ is normalized to unity
once the correct string tension is included in the relative
potential $V(p-q)$. This amounts to consider that the momentum $k$
distribution is a Dirac delta
\begin{equation}
(2 \pi)^3 \delta^3 \left(\trD{k} -\frac{\trD{p} +
\trD{q}}{2}\right) \ .
\end{equation}

If we summarize the above assumptions in a formula the loop vertex
that will be used is
\begin{equation}
\int\frac{d^4k}{(2\pi)^4}\!\left(\gamma^0
\frac{\lambda^a}{2}\right)\! S(k)\!\left(\gamma^0
\frac{\lambda^b}{2}\right)\! d^{abc} (2\pi)^3 \delta^3\!
\left(\trD{k} -\frac{\trD{p} + \trD{q}}{2}\right)\! .
\end{equation}

The equal-time approximation, which is standard in QMs, allows the
computation of the double vertex as a functional of the running
quark mass $m_k$. The quark mass will be computed
self-consistently in the next sections.

We decompose the fermion propagator in the usual particle and
antiparticle propagators
\begin{equation} \label{propagator}
\begin{split}
S(k) &= \frac{i}{\not k -m_k+i\epsilon} \\
&= \frac{i\Lambda^+(\trD{k})\beta }{k_0-E_k+i\epsilon} -
\frac{i\Lambda^-(\trD{k})\beta }{-k_0-E_k+i\epsilon}
\end{split}
\end{equation}
where the quark energy projectors are
\begin{equation} \label{projectors}
\begin{split}
\Lambda^+(\trD{k})&= \frac{1 + s_k \beta + c_k\,
\vers{k}\cdot\trD{\alpha}}{2}=\sum_s \spu{s}{k}\spudag{s}{k} \\
\Lambda^-(\trD{k})&= \frac{1 - s_k \beta - c_k\,
\vers{k}\cdot\trD{\alpha}}{2} =\sum_s \spv{s}{k}\spvdag{s}{k}
\end{split}
\end{equation}
and where $s_k=\sin\varphi_k =m_k/\sqrt{k^2+{m_k}^2}, \
c_k=\cos\varphi_k =k/\sqrt{k^2+{m_k}^2}$ and $\varphi_k$ is the
chiral angle, a convenient function for algebraic and numerical
computations.

The energy loop integral can be easily calculated,
\begin{equation}
\int\frac{dk^0}{2\pi}\frac{i}{k^0 \mp E \pm i\epsilon}\Lambda^\pm
\beta = \pm \Lambda^\pm \beta
\end{equation}
\begin{equation}
\begin{split}
\int\frac{dk^0}{2\pi}S(k) &= (\Lambda^+ -\Lambda^-)\beta \\
&= (s_k \beta + c_k\, \vers{k}\cdot\trD{\alpha})\beta
\end{split}
\end{equation}

Finally, using the Dirac delta distribution for the remaining
integrals over the three momentum $\trD{k}$, and summing in color
indices, we get the following effective vertex,
\begin{equation}
{\cal V}_{\mathrm{eff}} = {\cal C}\, \frac{\lambda^c}{2} \ (s_k -
c_k\, \vers{k}\cdot\trD{\gamma}) \ \Biggr|_{k=\frac{p+q}{2}} \ .
\label{vertex result}
\end{equation}
In the remainder of the paper we will always assume $k=(p+q)/2$.

Eq. (\ref{vertex result}) shows that the double vertex actually
solves the problem of matching chiral symmetry breaking and scalar
confinement. In the chiral limit of a vanishing quark mass, the
effective vertex ${\cal V}_{\mathrm{eff}}\rightarrow -{\cal C}\,
\lambda^c/2 \, \trD{k}\cdot\trD{\gamma}$ is proportional to the
$\gamma^\mu$ and is therefore chiral invariant as it should be,
whereas in the heavy quark limit, ${\cal V}_{\mathrm{eff}}
\rightarrow {\cal C} \lambda^c/2 $ is simply a scalar vertex. The
Gell-Mann matrix $\lambda^c$ provides the usual color vector
coupling as expected in a QM. We anticipate that the dynamical
generation of a quark mass will also generate a scalar coupling
for light quarks, and this results in an effective vertex,
including a chiral invariant vertex $\not \! p$ and the standard
scalar vertex $1$.

The dependence in the relative momentum must comply with the
linear confinement which is derived from the string picture,
\begin{equation}
\mathbf{V}_\varepsilon(\trD{x})= \frac{16}{3 {\cal C}^2}
V_\varepsilon(\trD{x})= \frac{16}{3 {\cal C}^2}\sigma |\trD{x}|
e^{-\varepsilon |\trD{x}|} \, ,
\end{equation}
where $\sigma \simeq 200$ MeV/Fm is the string constant and ${\cal
C}$ is the algebraic color factor defined in Eq. (\ref{color
factor}). The damping factor $\varepsilon$ regularizes the Fourier
transform,
\begin{equation}
\mathbf{V}_\varepsilon(\trD{p})= - \frac{16}{3{\cal C}^2}\,8\pi
\sigma \left(\frac{1}{(|\trD{p}|^2+\varepsilon^2)^2}
-\frac{4\varepsilon^2}{(|\trD{p}|^2+\varepsilon^2)^3}\right) \, ,
\end{equation}
and in the limit $\varepsilon \rightarrow 0$ we have
\begin{equation}
-i \, V_0( \trD{p}-\trD{q} )=-i \, \frac{-8\pi
\sigma}{|\trD{p}-\trD{q}|^4}\, .
\end{equation}

\section{Mass gap equation}
\label{sec:mass gap}

\begin{figure}
\includegraphics{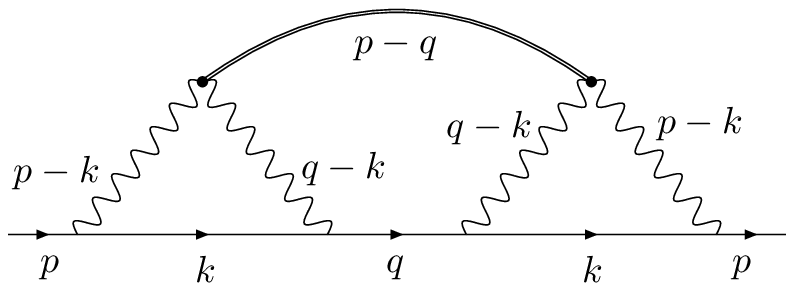}
\caption{ The self-energy term of the mass gap equation}
\label{massgapfig}
\end{figure}

We solve the mass gap equation using the Schwinger-Dyson formalism,
\begin{equation}
S^{-1} = {S_0}^{-1} - \Sigma  \ ,
\end{equation}
where the dressed propagator is defined in eq. (\ref{propagator})
and the free propagator has a similar definition with the quark
bare mass $m_0$. We want to determine the constituent quark mass
$m_p$ solving the self consistent mass gap equation. To compute
the self energy in this model, using the effective string
potential which couples to the quark line with double vertices, we
have to deal with the three loop diagram of Fig. \ref{massgapfig}.
Keeping in mind that the two double vertex loops are already
simplified. Technically other diagrams, with crossed gluon legs,
could exist but we assume that this is the dominant diagram.

The mass gap equation can be much simplified in the spin
formalism. Some useful relations we will use for this purpose are
\begin{equation} \label{simplrelations}
\begin{array}{l}
\spudag{s}{p}\, \spv{s'}{p}=
\trD{0}\cdot\left[\trD{\sigma}(i\sigma_2)\right]_{s s'} \\
\spudag{s}{p}\beta \spv{s'}{p}=
c_p\vers{p}\cdot\left[\trD{\sigma}(i\sigma_2)\right]_{s s'} \\
\spudag{s}{p}\alpha^i \spv{s'}{p}= -(\delta^{ij}-(1-
s_p)\hat{p}^i\hat{p}^j)\left[\sigma^j(i\sigma_2)\right]_{s s'} \\
\spudag{s}{p}\beta\alpha^i \spv{s'}{p}= -(s_p\delta^{ij}+(1-
s_p)\hat{p}^i\hat{p}^j)\left[\sigma^j(i\sigma_2)\right]_{s s'}
\end{array}
\end{equation}

With eqs. (\ref{propagator}), (\ref{projectors}) and
(\ref{simplrelations}) we arrive at the expected relation
\cite{bicudo.scalar}
\begin{equation}
\spub{s}{p}S^{-1}(p)\spv{s'}{p} =0\, ,
\end{equation}
implying that the propagator is diagonal in the
particle-antiparticle projection. The mass gap equation becomes
\begin{equation}
\spub{s}{p}S_0^{-1}(p)\spv{s'}{p} -
\spub{s}{p}\Sigma(p)\spv{s'}{p} =0\, .
\end{equation}

The free propagator term for a zero bare mass, $m_0=0$, is simply
\begin{equation} \label{mass gap free}
\spudag{s}{p}\beta (-i)\!\not \!p\ \spv{s'}{p} = i p\ s_p
\vers{p}\cdot\left[\trD{\sigma}(i\sigma_2)\right]_{s s'}
\end{equation}

The self-energy term is directly derived from the diagram of Fig.
\ref{massgapfig} using the double vertex loop ${\cal
V}_{\mathrm{eff}}$ already obtained in (\ref{vertex result}),
\begin{equation}
\begin{split}
\Sigma(p)=&\int \frac{d^3 q}{(2\pi)^3}
(s_k-c_k\vers{k}\cdot\trD{\gamma})
(s_q\beta+c_q\vers{q}\cdot\trD{\alpha})\beta \times \\
& \hspace{1cm} \times (s_k-c_k\vers{k}\cdot\trD{\gamma})\, {\cal
C}^2 \frac{3}{16} (-i) \mathbf{V}_\varepsilon(|\trD{p}-\trD{q}|) \\
=& \int\frac{d^3 q}{(2\pi)^3}\left( (s_k^2-c_k^2)s_q\beta - 2s_k
c_k s_q\, \vers{k}\cdot\trD{\alpha} + \right. \\
& + (s_k^2+c_k^2) c_q\, \vers{q}\cdot\trD{\alpha} +
2s_k c_k c_q\, \vers{k}\cdot\vers{q}\,\beta - \\
& \left. - 2c_k^2 c_q\, \vers{k}\cdot\vers{q}\,
\vers{k}\cdot\trD{\alpha}\right)\beta \,(-i)
V_\varepsilon(|\trD{p}-\trD{q}|) \ ,
\end{split}
\end{equation}
where we used the properties of the $\beta$ and $\alpha^i$
matrices. The result for the mass gap self-energy term is
\begin{equation} \label{mass gap selfenergy}
\begin{split}
& \spub{s}{p}\Sigma(p)\spv{s'}{p} = \\
& =\int\frac{d^3q}{(2\pi)^3} (-i)
V_\varepsilon\left(|\trD{p}-\trD{q}|\right)
\left[(s_k^2-c_k^2)s_q(-c_p \vers{p}) - \right. \\
& - 2s_k c_k s_q (-\vers{k}+(1-s_p)\vers{k}\cdot\vers{p}\
\vers{p}) + \\
& + c_q (-\vers{q}+(1-s_p)\vers{q}\cdot\vers{p}\ \vers{p}) + \\
& + 2s_k c_k c_q \vers{k}\cdot\vers{q}(-c_p \vers{p}) - \\
& \left.- 2c_k^2 c_q \vers{k}\cdot\vers{q}(-\vers{k}
+(1-s_p)\vers{k}\cdot\vers{p}\
\vers{p})\right]\cdot\left[\trD{\sigma}(i\sigma_2)\right]
\end{split}
\end{equation}
As we can see from (\ref{mass gap free}) and (\ref{mass gap
selfenergy}) both terms of the mass gap equation are proportional
to $\vers{p}\cdot\trD{\sigma}(i\sigma_2)$. Since the Pauli
matrices $\sigma$ are linearly independent, we can substitute
$\trD{\sigma}(i\sigma_2)$ by $\vers{p}$ and still have a mass gap
condition. With this simplification the mass gap equation reduces
to,
\begin{equation} \label{mass gap final}
\begin{split}
i p s_p - \int&\frac{d^3q}{(2\pi)^3}\left[ (c_k^2-s_k^2)s_q c_p +
2s_k c_k s_q s_p\vers{k}\cdot\vers{p} - \right. \\
& - c_q s_p\vers{q}\cdot\vers{p} - 2s_k c_k c_q c_p
\vers{k}\cdot\vers{q} + \\
& \left. + 2c_k^2 c_q s_p \vers{k}\cdot\vers{q}\
\vers{k}\cdot\vers{p} \right] i
V_\varepsilon\left(|\trD{p}-\trD{q}|\right)=0 \, .
\end{split}
\end{equation}
Notice that if we take the integrand and set $\trD{q}=\trD{p}$ we
will get $0\times V_\varepsilon(0)$. In section
\ref{sec:numerical} we will deal numerically with this IR
behavior.

\section{Numerical solution of the mass gap equation}
\label{sec:numerical}

The mass gap equation is a difficult non-linear integral equation,
that does not converge with the usual iterative methods. We
developed a method to solve the mass gap equation with a
differential equation, using a convergence parameter $\lambda$.
This parameter is the radius of a sphere centered in
$\trD{u}=\trD{p}-\trD{q}=0$ it allows us to separate the integral
into an integral inside the sphere and another outside of it,
\begin{equation}
\begin{split}
\int \frac{d^3u}{(2\pi)^3}\ f(\trD{p},\trD{u}) V_\varepsilon(u) =&
\int_\circ \frac{d^3u}{(2\pi)^3}\ f(\trD{p},\trD{u}) V_\varepsilon(u) \\
& + \int_{R^3-\circ} \frac{d^3u}{(2\pi)^3}\ f(\trD{p},\trD{u})
V_\varepsilon(u) \, .
\end{split}
\end{equation}
In our model $f(\trD{p},\trD{u})$ is the function dependent on the
chiral angle presented in eq. (\ref{mass gap final}).

Let us first focus on the integral inside the sphere, where we
have $u<\lambda$. Eventually we will take the limit where
$\lambda\rightarrow 0$ and this term will vanish. But for now we
will expand the function $f$ around $u=0$ and take only the first
non-vanishing term,

\begin{equation}
\begin{split}
\int_\circ \frac{d^3u}{(2\pi)^3}\ f(p,u,\omega)V_\varepsilon(u)
&\approx \int_\circ \frac{d^3u}{(2\pi)^3}\ \frac{\partial^2 f}
{\partial u^2}\Big|_{u=0} V_\varepsilon(u) \\
&\stackrel{\varepsilon \rightarrow 0}{=}
\frac{\lambda (-8\pi\sigma)}{8\pi^2} \int_{-1}^1 d\omega\
\frac{\partial^2 f} {\partial u^2}\Big|_{u=0} \, ,
\end{split}
\end{equation}
where $\omega$ is the cosine of the angle between $\trD{p}$ and
$\trD{u}$. For our particular model we have
\begin{equation}
\int_{-1}^1 d\omega\ \frac{\partial^2 f}{\partial u^2}|_{u=0} =
\frac{\sin(2\varphi_p) - 2p\cos(2\varphi_p)\varphi'_p +
p^2\varphi''_p}{3p^2}
\end{equation}

In what concerns the integral outside the sphere ($u>\lambda$) we
can take from the beginning $\varepsilon=0$ since this integral is
already regulated by $\lambda$. In this case we have
\begin{equation}
\begin{split}
\int_{R^3-\circ} \frac{d^3u}{(2\pi)^3}&\ f(\trD{p},\trD{u})V_0(u) = \\
& = \frac{-8\pi\sigma}{4\pi^2} \int_\lambda^\infty du \int^{-1}_1
d\omega\ \frac{1}{u^2} f(p,u,\omega) \, .
\end{split}
\end{equation}

Placing the two terms in the mass gap equation we finally get the
equation that we can iterate to find the solution,

\begin{widetext}
\begin{equation}
3 p\ s_p + \frac{\sigma}{\pi} \left[ \lambda\ (\varphi''_p -
\frac{2}{p}\cos(2\varphi_p)\varphi'_p +
\frac{1}{p^2}\sin(2\varphi_p)) + 6 \int_\lambda^\infty du
\int^{-1}_1 d\omega\ \frac{1}{u^2} f(p,u,\omega) \right]=0
\label{numerical equation}
\end{equation}

\end{widetext}

Our technique consists in starting with a large infrared cutoff
$\lambda$, where the integral term of Eq. (\ref{numerical
equation}) is negligible. In this case Eq. (\ref{numerical
equation}) for the chiral angle $\varphi_p$ becomes essentially a
differential equation which can be solved with the standard
shooting method \cite{bicudo.tese}. It turns out that this
equation possesses several solutions, and we specialize in the
larger one, with no nodes, that corresponds to the stable vacuum
\cite{bicudo.replica}. Then one decreases step by step the
$\lambda$ parameter, using as an initial guess for the evaluation
of the integral the $\varphi_p$ determined for the previous value
of $\lambda$. In this way the integral is a simple function of the
momentum $p$ and we again have to solve a non-homogeneous
differential equation. Eventually we are able to solve the mass
gap equation for a $\lambda$ parameter which is much smaller than
the scale of the interaction. Finally we extrapolate the set of
obtained $\varphi_p$ to the limit of $\lambda \rightarrow 0$.
\begin{figure}
\includegraphics{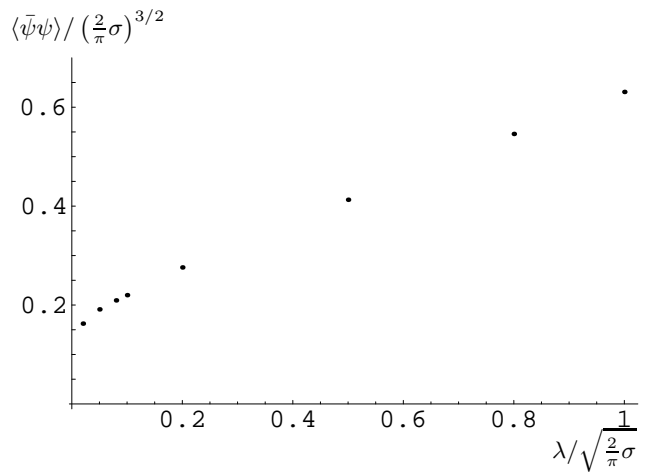}

\begin{picture}(0,0)(0,0)
\put(-120,163){$\langle \bar \psi \psi \rangle
/\left(\frac{2}{\pi}\sigma\right)^{3/2} $}
\put(85,3){$\lambda/\sqrt{\frac{2}{\pi}\sigma}$}
\end{picture}
\caption{ Testing the convergence of the numerical method with the
quark condensate $\langle \bar \psi \psi \rangle$.}

\label{testing in condensate}
\end{figure}
We test the convergence of the method computing the quark
condensate $\langle \bar \psi \psi \rangle $,
\begin{equation}
\langle \bar \psi \psi \rangle = 6 \int \frac{d^3 p}{(2 \pi)^3}
s_p \, .
\end{equation}
The evolution of the solution as a function of the infrared
parameter $\lambda$ is clearer when we display the quark
condensate, see Fig. \ref{testing in condensate}.

\begin{figure}
\includegraphics{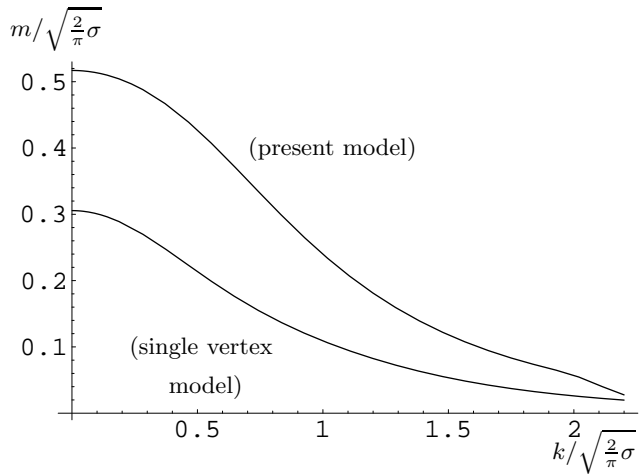}

\begin{picture}(0,0)(0,0)
\put(-30,120){(present model)}
\put(-75,45){(single vertex}
\put(-60,30){model)} \put(85,5){$k/\sqrt{\frac{2}{\pi}\sigma}$}
\put(-120,165){$m/\sqrt{\frac{2}{\pi}\sigma}$}
\end{picture}

\caption{ The $m_k$ solutions of the mass gap equation in units of
$\sqrt{{2\over \pi}\sigma}$.} \label{massgapsolved}
\end{figure}

The solution of the mass gap equation is presented in Fig.
\ref{massgapsolved}, where we compare it with the single vertex
model solution.

\section{Results and conclusion}
\label{sec:conclusion}

\begin{figure}
\includegraphics{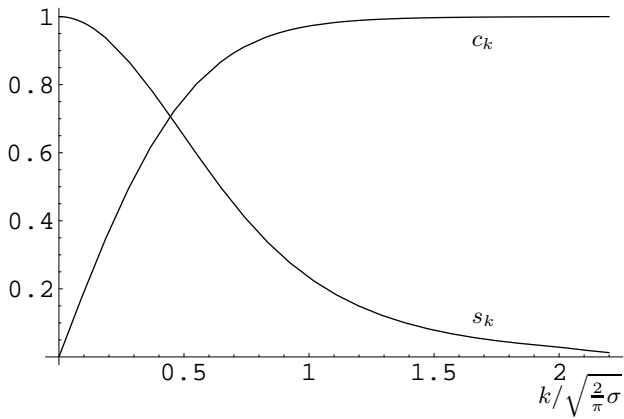}

\begin{picture}(0,0)(0,0)
\put(60,35){$s_k$} \put(60,140){$c_k$}
\put(85,5){$k/\sqrt{\frac{2}{\pi}\sigma}$}
\end{picture}

\caption{Form factors for the scalar vertex ($s_k$) and for the
remaining chiral invariant vertex ($c_k$)} \label{form factors}
\end{figure}

In this paper we build a QM for the coupling of quark to a scalar
string. The quark confining interaction has a single parameter
$\sigma$. This QM matches the apparently conflicting vector
coupling of QCD with a scalar confinement. Our model can be
interpreted as a double vertex that couples the quark to the
string, or alternatively as a $\gamma_0 S(k)\gamma_0$ vertex.
Either way this vertex decomposes in the sum of a scalar vertex
$1$ and a chiral invariant $\vers{k}\cdot\trD{\gamma}$ vertex
weighed by simple functions of the dynamical quark mass. In the
chiral limit the scalar vertex vanishes, while in the heavy quark
limit the confining potential is essentially scalar. Our results
for the weighing factors of the scalar vertex and the remaining
chiral invariant vertex are shown in Fig. \ref{form factors}.

We solve the mass gap equation for the dynamical generation of the
quark mass with the S$\chi$SB, and we indeed generate the
constituent quark mass. We show that S$\chi$SB not only generates
a quark mass, but generates also a scalar vertex for the
confinement. The results are encouraging because the quark
condensate indeed increases when compared with the simpler one
vertex vector confining potential.

It is clear that the next step of this work, will consist in
adding the shorter range one gluon exchange potential to the
confining potential. The resulting model will have two parameters,
one for the short range potential, and another one for the
confining potential. These parameters will be determined in the
fit of the hadron spectrum. In what concerns the mass gap
equation, we expect that this will further enhance the quark
condensate, possibly up to the expected $- (230 \, MeV)^3$.

\acknowledgements

Pedro Bicudo thanks discussions on scalar or string confinement
with Alfredo Henriques, Jack Paton, Franz Gross, Dieter Gromes,
Nora Brambilla, Jean-Fran\c{c}ois Lagae, Mike Pichowsky,
Misha Polikarpov, Dimitri Diakonov and Alexei Nefediev.

The work of G. M. Marques is supported by Funda\c c\~ao para a
Ci\^encia e a Tecnologia under the grant SFRH/BD/984/2000.


\end{document}